\documentclass[aps,prc,preprint,showpacs,superscriptaddress]{revtex4}  
\usepackage{graphicx}
\usepackage{amsmath}
\begin{document}
\newcommand {\nc} {\newcommand}
\nc {\beq} {\begin{eqnarray}} 
\nc {\eol} {\nonumber \\} 
\nc {\eeq} {\end{eqnarray}} 
\nc {\eeqn} [1] {\label{#1} \end{eqnarray}}   
\nc {\eoln} [1] {\label{#1} \\} 
\nc {\ve} [1] {\mbox{\boldmath $#1$}}
\nc {\rref} [1] {(\ref{#1})} 
\nc {\Eq} [1] {Eq.~(\ref{#1})} 
\nc {\la} {\mbox{$\langle$}}
\nc {\ra} {\mbox{$\rangle$}}
\nc {\dd} {\mbox{${\rm d}$}}
\nc {\cM} {\mathcal{M}} 
\nc {\cY} {\mathcal{Y}} 
\nc {\dem} {\mbox{$\frac{1}{2}$}}
\nc {\ut} {\mbox{$\frac{1}{3}$}} 
\nc {\qt} {\mbox{$\frac{4}{3}$}} 
\nc {\Li} {\mbox{$^6\mathrm{Li}$}}
\nc {\M} {\mbox{$\mathcal{M}$}}

\title{Influence of orthogonalization procedure on astrophysical S-factor for the direct $\alpha+d$ $\rightarrow$ $^6$Li + $\gamma $  capture process in a three-body model}

\author{E.M. Tursunov}

\address{Institute of Nuclear Physics,
Academy of Sciences, 100214, Ulugbek, Tashkent, Uzbekistan\\
tursune@inp.uz}

\author{A.S. Kadyrov}

\address{Curtin Institute for Computation and Department of Physics and Astronomy, Curtin University, GPO Box U1987, Perth, WA 6845, Australia\\
a.kadyrov@curtin.edu.au}

\begin{abstract}
The astrophysical S-factor for the direct  $ \alpha(d,\gamma)^{6}{\rm Li}$ capture reaction is calculated in a three-body model based on  the hyperspherical Lagrange-mesh method. A sensitivity of the E1 and E2 astrophysical S-factors to the orthogonalization method of Pauli forbidden states  in the three-body system is studied. 
It is found that the method of orthogonalising pseudopotentials (OPP) yields larger isotriplet ($T=1$) components than the supersymmetric transformation (SUSY) procedure. The E1 astrophysical S-factor shows the same energy dependence in both cases, but strongly different absolute values. At the same time,   the E2 S-factor does not depend on the orthogonalization procedure. As a result, the  OPP method yields a very good description of the direct data of the  LUNA collaboration at low energies, while the SUSY transformation strongly underestimates the LUNA data.  
\end{abstract}

\maketitle

\section{Introduction}

\par A consistent realistic estimation of the primordial abundance ratio  $^{6}$Li/ $^{7}$Li
of the lithium isotopes is one of the open problems in nuclear astrophysics.
 For this  ratio the BBN model \cite{serp04} yields a value about
three orders of magnitude less than the astrophysical data \cite{asp06}.  
One input  parameter for the estimation of the abundance ratio is the reaction rates of the direct   radiative
capture process
\begin{eqnarray} \label{1}
\alpha+d\rightarrow {\rm ^6Li}+\gamma
\end{eqnarray}
at low energies within the range $30 \le E_{\rm cm} \le 400$ keV
\cite{serp04}.  

The reaction rate evaluations are based on the straightforward calculations starting from the astrophysical S-factor.
The data set of the LUNA collaboration available at 
 energies E=94 keV and E=134 keV \cite{luna14} was
recently renewed with additional data at E=80 keV and E=120 keV  \cite{luna17}.  The results of the LUNA collaboration for the reaction rates turn out to be even lower than previously reported. 
 This further increases the discrepancy between prediction of the  BBN model
 and the astronomical observations for the  primordial abundance of the  $^6$Li element in the Universe \cite{luna17}.

From a theoretical point of view,  an  important step toward the solution of the lithium abundance problem has been taken within the three-body model \cite{tur16,bt18,tur18}. 
As was shown  in Ref. \cite{bt18} in detail,  the so-called exact mass prescription, used in the literature for the estimation of the E1 astrophysical S-factor during a  long period \cite{type97,mukh11,tur15,noll01,TBL91},  has no microscopic background at all.  Some models even neglect this important contribution to the S-factor \cite{desc98,lang86}. As was shown within the frame of three-body model based on the hyperspherical Lagrange mesh method \cite{bt18,tur18}, the E1 S-factor is dominant in the low  energy region $E<100$ keV, while the E2 S-factor is mostly important at higher energies.  As a result, the new data of the LUNA collaboration for the astrophysical S-factor at low energies have been reproduced with a good accuracy.  The estimated $^6$Li/H abundance ratio of $(0.67 \pm 0.01)\times 10^{-14}$ was in a very good agreement with the experimental value of $(0.80 \pm 0.18)\times 10^{-14}$ from the LUNA collaboration.  

On the other hand, the final three-body $\alpha+p+n$ hyperspherical wave function of $^6$Li was calculated with the Voronchev et al. $\alpha N$ -potentials \cite{vor95}  with a forbidden state in the S-wave.  The forbidden states in the three-body system have been treated within the method of orthogonalising pseudopotentials (OPP) \cite{OPP}. The wave function  has a small isotriplet component of about 0.5 percent. This important  isotriplet part of the three-body wave function \cite{tur16,bt18,tur18} enables one to estimate the forbidden E1 S-factor in a consistent way without using any exact mass prescription.  
An important question is, how sensitive are the results for the estimated astrophysical S-factor on the projecting method used in the variational calculations of the three-body wave function. The aim of  present study is to  answer this important question. To this end  we estimate the astrophysical S-factor with the three-body wave function of the $^6$Li ground state, calculated using the supersymmetric transformation (SUSY) method \cite{baye87} and compare with the results of the OPP approach.

\section{Theoretical model}
 
The cross sections of the radiative capture process reads as
\begin{align}
\sigma_{E}(\lambda)=& \sum_{J_i T_i \pi_i}\sum_{J_f T_f
\pi_f}\sum_{\Omega \lambda}\frac{(2J_f+1)} {\left [I_1
\right]\left[I_2\right]} \frac{32 \pi^2 (\lambda+1)}{\hbar \lambda
\left( \left[ \lambda \right]!! \right)^2} k_{\gamma}^{2 \lambda+1}
\nonumber \\ &\times \sum_{l_\omega I_\omega}
 \frac{1}{ k_\omega^2 v_\omega}\mid
 \langle \Psi^{J_f T_f \pi_f}
\|M_\lambda^\Omega\|
\Psi_{l_\omega I_\omega}^{J_i T_i \pi_i}
\rangle \mid^2,
\end{align}
where $\Omega=$E  or M (electric or magnetic transition), $\omega$
denotes the entrance channel, $k_{\omega}$, $v_\omega$,  $I_\omega$
are the wave number, velocity of the $\alpha-d$ relative motion and
the spin of the entrance channel, respectively, $J_f$, $T_f$,
$\pi_f$ ($J_i$, $T_i$, $\pi_i$) are the spin, isospin and parity of
the final (initial) state, $I_1$, $I_2$ are channel spins,
$k_{\gamma}=E_\gamma / \hbar c$ is the wave number of the photon
corresponding to the energy $E_\gamma=E_{\rm th}+E$ with the
threshold energy $E_{\rm th}=1.474$ MeV. The wave functions
$\Psi_{l_\omega I_\omega}^{J_i T_i \pi_i} $ and $\Psi^{J_f T_f
\pi_f} $ represent  the initial and final states, respectively. The
reduced matrix elements of the transition operators are evaluated between the initial and final
states. We also use short-hand notations $[I]=2I+1$ and
$[\lambda]!!=(2\lambda+1)!!$.  Details of the matrix-element calculations have been given in Ref.\cite{tur16}.

The astrophysical $S$-factor of the process is expressed in terms of  the
cross section as \cite{Fowler}
\begin{eqnarray}
S(E)=E \, \sigma_E(\lambda) \exp(2 \pi \eta),
\end{eqnarray}
where  $\eta$ is the Coulomb parameter.

\section{Numerical results}
 
Calculations of the cross section and astrophysical S-factor have
been performed under the same conditions as in Refs.\cite{bt18,tur18}. The
radial wave function of the deuteron is the solution of the
bound-state Schr{\"o}dinger equation with the central Minnesota
potential $V_{NN}$ \cite{thom77,RT70} with $\hbar^2/2
m_N=20.7343$ MeV fm$^2$. The Schr{\"o}dinger equation is solved
using a highly accurate Lagrange-Laguerre mesh method \cite{baye15}.
It yields $E_d$=-2.202 MeV for the deuteron ground-state energy with
the number of mesh points $N=40$ and a scaling parameter $h_d=0.40$.

The scattering wave function of the $\alpha-d$ relative motion is
calculated with a deep potential of Dubovichenko \cite{dub94} with a
small modification in the $S$-wave \cite{tur15}:
$V_d^{(S)}(R)=-92.44 \exp(- 0.25 R^2) $ MeV. The potential
parameters in the $^3P_0$, $^3P_1$, $^3P_2$  and $^3D_1$, $^3D_2$,
$^3D_3$ partial waves are the same as in Ref. \cite{dub94}. The
potential contains additional states in the $S$- and $P$-waves
forbidden by the Pauli principle. The above modification of the
S-wave potential reproduces the empirical value $C_{\alpha d}=2.31$
fm$^{-1/2}$ of the asymptotic normalization coefficient (ANC) of the
$^6$Li(1+) ground state derived from $\alpha-d$ elastic scattering
data \cite{blok93}.

The final $^6$Li(1+) ground-state wave function was calculated using
the hyperspherical Lagrange-mesh method \cite{tur07}
with the same Minnesota NN-potential. For the $\alpha-N$ nuclear
interaction the potential of Voronchev {\em et al.}  
\cite{vor95} was
employed, which contains a deep Pauli-forbidden state in the
$S$-wave. The potential was slightly renormalized by a scaling
factors 1.014  to reproduce the
experimental binding energy of $E_b$=3.70 MeV. The Coulomb interaction
between $\alpha$ and proton is taken as $2e^2\, \mathrm{erf}(0.83\,
R)/R$ \cite{RT70}. The coupled hyperradial equations are solved with
the Lagrange-mesh method \cite{baye15}. 
The hypermomentum expansion includes terms up to a large value of $K_{\rm max} $,
 which ensures a good convergence of the energy and of the $T = 1$ component of
$^6$Li.  In Refs.\cite{bt18,tur18} the OPP method was used for the treatment of the Pauli forbidden states in the three-body model. Here we also examine the SUSY transformation \cite{baye87} of the initial  $\alpha-N$ nuclear
interaction potential. This operation yields a shallow potential which gives the same phase shift, 
but  removes unphysical forbiddden state from the S-wave $\alpha+N$ spectrum.  

Firstly, the energy convergence in the three-body $\alpha+p+n$ system for the SUSY and OPP methods shows the same behavior.
The energy of the $^6$Li ground state $E=-3.70$  MeV converges already at a maximal hypermomentum $K_{max}=24$ in the both cases.
However, the structure of the $^6$Li g.s. wave function in these two versions of the projection yields different pictures.
The important isotriplet  ($T=1$)  component of  the $^6$Li g.s. wave function used in the OPP method has a norm square of about 5.27$\times$10$^{-3}$, while in the case of the SUSY method it is 1.10$\times$10$^{-4}$.  As we noted above, the important isotriplet
 component of the final $^6$Li ground state 
 is responsible for the E1 astrophysical S-factor in the $ \alpha(d,
\gamma)^{6}{\rm Li}$ direct capture reaction. Therefore, the above  difference should yield the same 
effect for the  E1 S-factor. Additionally, it is important to check, whether the energy dependence of the E1 S-factor is the same in both cases.
 
\begin{figure}[htb]
\centerline{\includegraphics[width=10.7cm]{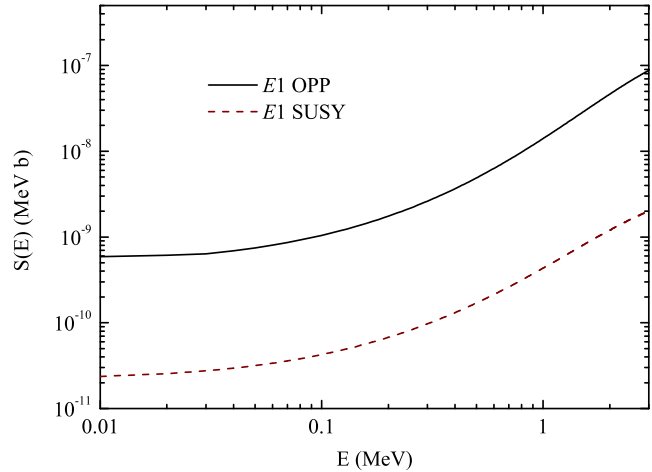}} \vspace*{8pt}
\caption{Astrophysical E1 S-factor of the direct $ \alpha(d,
\gamma)^{6}{\rm Li}$ capture process. 
\label{f1}}
\end{figure}

\begin{figure}[htb]
\centerline{\includegraphics[width=10.7cm]{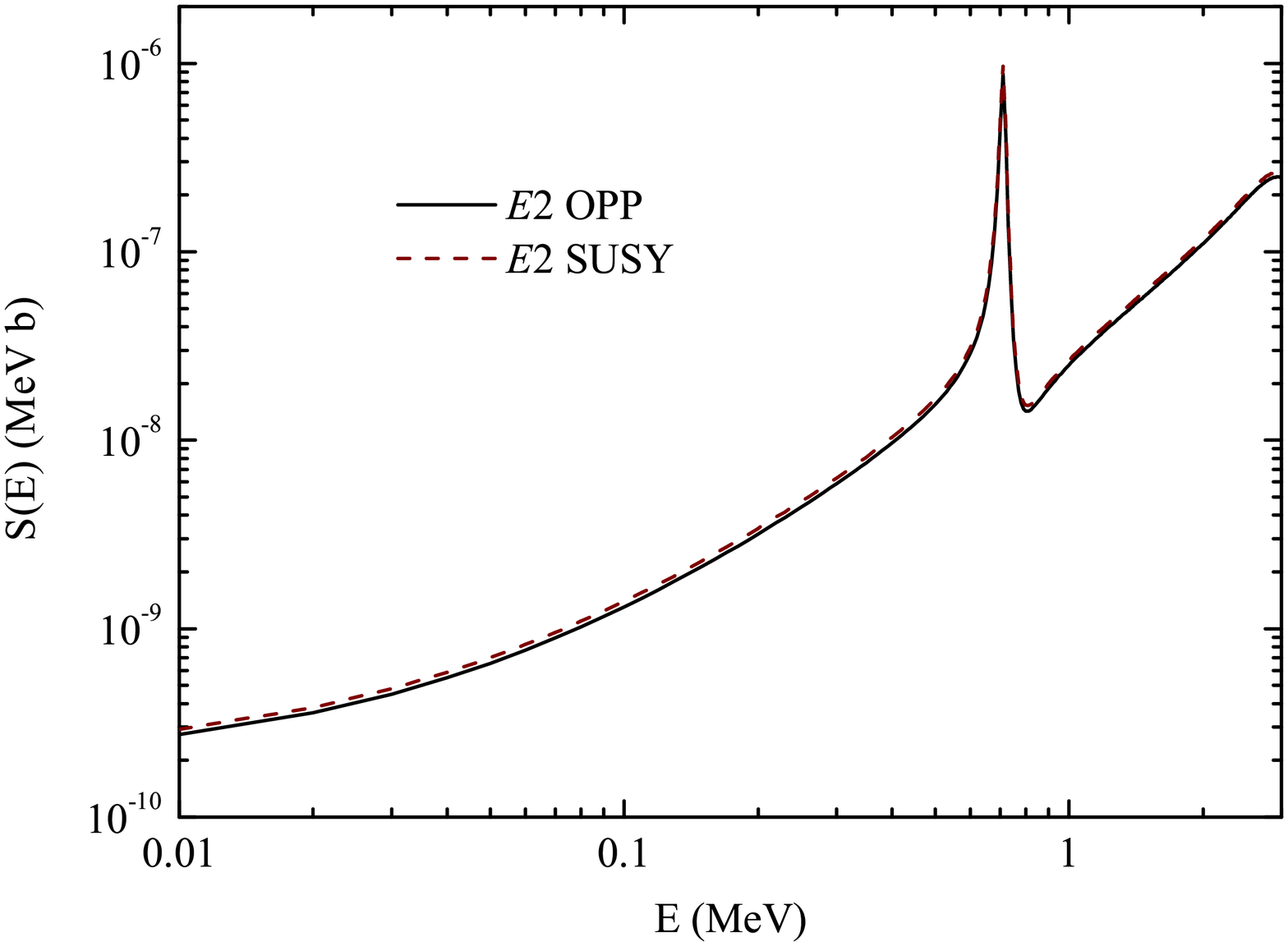}} \vspace*{8pt}
\caption{Astrophysical E2 S-factor of the direct $ \alpha(d,
\gamma)^{6}{\rm Li}$ capture process. 
\label{f2}}
\end{figure}

\begin{figure}[htb]
\centerline{\includegraphics[width=10.7cm]{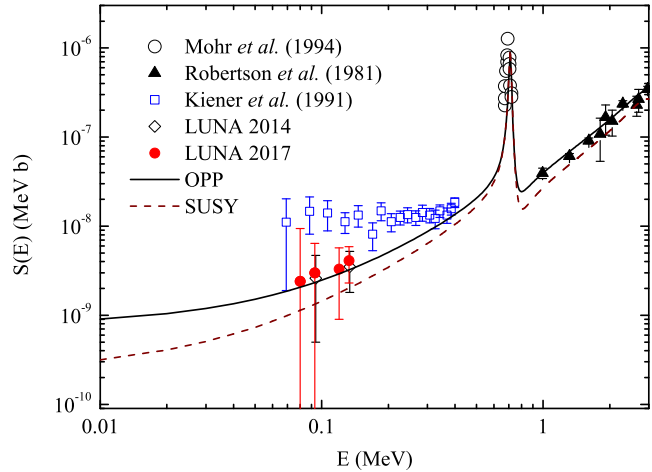}} \vspace*{8pt}
\caption{Astrophysical S-factor of the direct $ \alpha(d,
\gamma)^{6}{\rm Li}$ capture process.
\label{f3}}
\end{figure}

In Fig. \ref{f1} we show the E1 astrophysical S-factor of the direct   $ \alpha(d,
\gamma)^{6}{\rm Li}$ capture process  estimated with the OPP and SUSY three-body wave functions. As we can see from the 
figure, the two methods yield the same energy behavior. However, the SUSY method yields too small S-factor in the entire energy region, more than one order of magnitude smaller than the OPP method. This result indicates that the E1 astrophysical S-factor is highly sensitive to the orthogonalization method. 
A similar effect was found in the beta decay of the $^6$He halo nucleus \cite{tur06,tur06a} and M1-transition of the $^6$Li(0+) isobar-analog state to the $\alpha+d$ continuum \cite{tur07}.  In fact, the OPP method yields scattering and bound state wave functions with a node at short distances, while this nodal behavior disappears in the  SUSY method.  In the present study, a nodal behavior of the S-wave $\alpha+N$ wave function yields  a strong contribution to the important isotriplet component  of the total $^6$Li ground state wave function.  

The E2 astrophysical S-factor is displayed in Fig. \ref{f2}. As can be seen, the OPP and SUSY methods give the same theoretical estimations. This means that the  E2  S-factor is not sensitive to the orthogonalization procedure in the wave function of the $^6$Li ground state.

The total theoretical astrophysical S-factor for the process is shown in Fig. \ref{f3}  in comparison with the direct data of the LUNA collaboration \cite{luna14,luna17} and old data from Refs.\cite{rob81,kien91,mohr94}. Due to a strong effect of the orthogonalization method we have a big difference in the SUSY and OPP results. While the OPP method yields a good description of the direct data of the LUNA collaboration, the SUSY transformation gives a strong underestimation. 

\section{Conclusion}
A sensitivity of the theoretical astrophysical S-factor for the direct  $ \alpha(d,\gamma)^{6}{\rm Li}$ capture reaction  to the orthogonalization procedure has been examined within the hyperspherical Lagrange-mesh method. It was found that the E1 astrophysical S-factor is very sensitive to the orthogonalization method, however the E2 S-factor does not depend on the orthogonalization procedure.  As a result, the OPP method yields a very good description of the direct data of the  LUNA collaboration at low energies, 
while the SUSY transformation significantly underestimates the LUNA data.  On the other hand, both methods show the same energy dependence for the E1 S-factor. 

The authors thank D. Baye and P. Descouvemont for very useful discussions of the presented results.

\end{document}